\documentclass[graybox]{svmult}

\usepackage{mathptmx}       % selects Times Roman as basic font
\usepackage{helvet}         % selects Helvetica as sans-serif font
\usepackage{courier}        % selects Courier as typewriter font
\usepackage{makeidx}         % allows index generation
\usepackage{graphicx}        % standard LaTeX graphics tool
\usepackage{multicol}        % used for the two-column index
\usepackage[bottom]{footmisc}% places footnotes at page bottom

\usepackage{amssymb}

\makeindex             % used for the subject index
\begin{document}

\title*{Motion in alternative theories of gravity\\
{\small (lecture given at the \textit{School on Mass}, Orl\'eans, France, 23--25 June 2008)}}
\author{Gilles Esposito-Far\`ese}
\institute{Gilles Esposito-Far\`ese \at
${\mathcal{G}}{\mathbb{R}}\varepsilon{\mathbb{C}}{\mathcal{O}}$,
Institut d'Astrophysique de Paris, UMR 7095-CNRS,
Universit\'e Pierre et Marie Curie-Paris~6,
98bis boulevard Arago, F-75014 Paris, France,
\email{gef@iap.fr}}
\maketitle

\abstract*{Although general relativity (GR) passes all present
experimental tests with flying colors, it remains important to
study alternative theories of gravity for several theoretical and
phenomenological reasons that we recall in these lecture notes.
The various possible ways of modifying GR are presented, and we
notably show that the motion of massive bodies may be changed
even if one assumes that matter is minimally coupled to the
metric as in GR. This is illustrated with the particular case of
scalar-tensor theories of gravity, whose Fokker action is
discussed, and we also mention the consequences of the no-hair
theorem on the motion of black holes. The finite size of the
bodies modifies their motion with respect to pointlike particles,
and we give a simple argument showing that the corresponding
effects are generically much larger in alternative theories than
in GR. We also discuss possible modifications of Newtonian
dynamics (MOND) at large distances, which have been proposed to
avoid the dark matter hypothesis. We underline that all the
previous classes of alternatives to GR may {\it a priori\/} be
used to predict such a phenomenology, but that they generically
involve several theoretical and experimental difficulties.}

\abstract{Although general relativity (GR) passes all present
experimental tests with flying colors, it remains important to
study alternative theories of gravity for several theoretical and
phenomenological reasons that we recall in these lecture notes.
The various possible ways of modifying GR are presented, and we
notably show that the motion of massive bodies may be changed
even if one assumes that matter is minimally coupled to the
metric as in GR. This is illustrated with the particular case of
scalar-tensor theories of gravity, whose Fokker action is
discussed, and we also mention the consequences of the no-hair
theorem on the motion of black holes. The finite size of the
bodies modifies their motion with respect to pointlike particles,
and we give a simple argument showing that the corresponding
effects are generically much larger in alternative theories than
in GR. We also discuss possible modifications of Newtonian
dynamics (MOND) at large distances, which have been proposed to
avoid the dark matter hypothesis. We underline that all the
previous classes of alternatives to GR may {\it a priori\/} be
used to predict such a phenomenology, but that they generically
involve several theoretical and experimental difficulties.}

\section{Introduction}
\label{sec:1}
Since general relativity (GR) is superbly consistent with all
precision experimental tests ---~as we will see below several
examples, one may naturally ask the question: Why should we
consider alternative theories of gravity? The reason is actually
threefold. First, it is quite instructive to contrast GR's
predictions with those of alternative models in order to
understand better which features of the theory have been
experimentally tested, and what new observations may allow us to
test the remaining features. Second, theoretical attempts at
quantizing gravity or unifying it with other interactions
generically predict the existence of partners to the graviton,
i.e., extra fields contributing to the gravitational force. This
is notably the case in all extra-dimensional (Kaluza-Klein)
theories, where the components $g_{ab}$ of the metric tensor
[where $a$ and $b$ belong to the $D-4$ extra dimensions] behave
as $(D-4)(D-3)/2$ scalar fields (called moduli) in four
dimensions. Independently of such moduli, supersymmetry (notably
needed in string theory) also implies the existence of several
fields in the graviton supermultiplet, in particular another
scalar called the dilaton. The third reason why it remains
important to study alternative theories of gravity is the
existence of several puzzling experimental issues. Cosmological
data are notably consistent with a Universe filled with about
72\% of ``dark energy'' (a fluid with negative pressure opposite
to its energy density) and $24\%$ of ``dark matter'' (a fluid
with negligible pressure and vanishingly small interaction with
ordinary matter and itself) \cite{Spergel:2003cb,Spergel:2006hy}.
Another strange phenomenon is the anomalous extra acceleration
towards the Sun that the two Pioneer spacecrafts have undergone
beyond 30 astronomical units
\cite{Anderson:2001sg,Nieto:2003rq,Turyshev:2005ej} (see also
\cite{Lammerzahl:2006ex}). Although such issues do not threaten
directly GR itself, since they may be explained by the existence
of unknown ``dark'' fluids (or by yet unmodelled sources of noise
in the case of the Pioneer anomaly), they may nevertheless be a
hint that something needs to be changed in the gravitational law
at large distances.

Since many theoretical and experimental physicists devised their
own gravity models, the field of alternative theories is much too
wide for the present lecture notes. Detailed reviews may be found
in Refs.~\cite{Willbook,WillLivRev,UzanConst,DamourPDG}. We shall
focus here on the particular case of scalar-tensor theories and
some of their generalizations. Before introducing them, let us
recall that GR is based on two independent hypotheses, which can
be most conveniently described by writing its action
\begin{equation}\label{eq:SGR}
S = \underbrace{\frac{c^3}{16\pi G}
\int{d^4 x\sqrt{-g}\, R}}_{\rm Einstein-Hilbert}
+\underbrace{\vphantom{\int}S_{\rm matter}[{\rm matter},
g_{\mu\nu}]}_{\rm metric\ coupling},
\end{equation}
where $g$ denotes the determinant of the metric $g_{\mu\nu}$, $R$
its scalar curvature, and we use the sign conventions of
Ref.~\cite{MTW}, notably the mostly-plus signature. The first
assumption of GR is that matter fields are universally and
minimally coupled to one single metric tensor $g_{\mu\nu}$. This
ensures the ``Einstein equivalence principle'', whose
consequences will be summarized in Sec.~\ref{sec:2} below. The
second hypothesis of GR is that this metric $g_{\mu\nu}$
propagates as a pure spin-2 field, i.e., that its kinetic term is
given by the Einstein-Hilbert action. The core of the present
lecture notes, Secs.~\ref{sec:3} to \ref{sec:6}, will be devoted
to the observational consequences of other possible kinetic
terms.

\section{Modifying the matter action}
\label{sec:2}
In the above action (\ref{eq:SGR}), square brackets in $S_{\rm
matter}[{\rm matter}, g_{\mu\nu}]$ mean a functional dependence
on the fields, i.e., it also depends on their first derivatives.
For instance, the action of a point particle,
\begin{equation}\label{eq:Spp}
S_{\rm point\ particle} = - \int m c\, ds =
-\int m c\sqrt{-g_{\mu\nu}(x)\,v^\mu v^\nu}\, dt,
\end{equation}
depends not only on its spacetime position $x$ but also on its
4-velocity $v^\mu \equiv dx^\mu/dt$. Since the matter action
defines the \textit{motion} of matter in a given metric
$g_{\mu\nu}$, it is \textit{a priori} what needs to be modified
with respect to GR in order to predict different trajectories.
This idea has been studied in depth by Milgrom in
\cite{Milgrom:1992hr,Milgrom:1998sy}, where he assumed that the
action of a point particle could also depend on its acceleration
and even higher time-derivatives: $S_{\rm pp}(\vec{x}, \vec{v},
\vec{a}, \dot{\vec{a}}, \dots)$. However, any modification with
respect to action (\ref{eq:Spp}) is tightly constrained
experimentally for usual accelerations, notably by high-precision
tests of special relativity. On the other hand, physics may
happen to differ for tiny accelerations, much smaller than the
Earth's gravitational attraction. In such a case, the
mathematical consistency of the theory may be invoked to restrict
the space of allowed theories. A theorem derived by Ostrogradski
in 1850 \cite{Ostrogradski,Woodard:2006nt} shows notably that the
Hamiltonian is generically unbounded from below if $S_{\rm
pp}(\vec{x}, \vec{v}, \vec{a}, \dots, d^n \vec{x}/dt^n)$ depends
on a \textit{finite} number of time-derivatives, and therefore
that the theory is unstable. A possible solution would thus be to
consider \textit{nonlocal} theories, depending on a infinite
number of time derivatives. This is actually what Milgrom found
to be necessary in order to recover the Newtonian limit and
satisfy Galileo invariance. Although nonlocal theories are worth
studying, and are actually obtained as effective models of string
theory, their phenomenology is quite difficult to analyze, and we
will not consider them any longer in the present lecture notes.
General discussions and specific models may be found for instance
in \cite{Simon:1990ic,Milgrom:1992hr,Soussa:2003vv,Deser:2007jk,
Deffayet:2009ca}.

Another possible modification of the matter action $S_{\rm
matter}[{\rm matter}, g_{\mu\nu}]$ is actually predicted by
string theory: Different matter fields are coupled to different
metric tensors, and the action takes thus the form $S_{\rm
matter}[{\rm matter}_{\vphantom{\mu}}^{(i)}, g_{\mu\nu}^{(i)}]$.
In other words, two different bodies \textit{a priori} do not
feel the same geometry, and their accelerations may differ both
in norm and direction. However, the universality of free fall is
extremely well tested experimentally, as well as the three other
observational consequences of a metric coupling $S_{\rm
matter}[{\rm matter}, g_{\mu\nu}]$, that we will recall below.
The conclusion is that string theory must actually show that the
different metrics $g_{\mu\nu}^{(i)}$ are almost equal to each
other. One possible reason is that their differences may be
mediated by massive fields, and would become thus exponentially
small at large enough distances. But even in presence of massless
fields contributing to the difference between the various
$g_{\mu\nu}^{(i)}$, a generic mechanism has been shown to attract
the theory towards GR during the cosmological expansion of the
Universe \cite{DamourPolyakov,DamourVilenkin,DamourVeneziano}.

Let us now recall the four observational consequences of a metric
coupling $S_{\rm matter}[{\rm matter}, g_{\mu\nu}]$, as well as
their best experimental verifications. If all matter fields feel
the same metric $g_{\mu\nu}$, it is possible to define a ``Fermi
coordinate system'' along any worldline, such that the metric
takes the diagonal form ${\rm diag}(-1,1,1,1)$ and its first
derivatives vanish. In other words, up to small tidal effects
proportional to the spatial distance to the worldline, everything
behaves as in special relativity. This is the mathematically
well-defined notion of a freely-falling elevator. The effacement
of gravity in this coordinate system implies that (i)~all
coupling constants and mass scales of the Standard Model of
particle physics are indeed space and time independent. One of
the best experimental confirmations is the time-independence of
the fine-structure constant, $|\dot\alpha/\alpha| < 7\times
10^{-17} {\rm yr}^{-1}$, six orders of magnitude smaller than the
inverse age of the Universe \cite{Shlyakhter,DamourDyson,Fujii}.
A second consequence of the validity of special relativity within
the freely-falling elevator is that (ii)~local
(non-gravitational) experiments are Lorentz invariant. The
isotropy of space has notably been tested at the $10^{-27}$ level
in \cite{Prestage,Lamoreaux,Chupp}. The third consequence of a
metric coupling is the very existence of this freely falling
elevator where gravity is effaced, i.e., (iii)~the universality
of free fall: (non self-gravitating) bodies fall with the same
acceleration in an external gravitational field. This has been
tested at a few parts in $10^{13}$ both in laboratory experiments
\cite{Baessler:1999iv,Adelberger}, and by studying the relative
acceleration of the Earth and the Moon towards the Sun
\cite{Williams:2004qb}. The fourth consequence of a metric
coupling is (iv)~the universality of gravitational redshift. It
may be understood intuitively by invoking the equivalence between
the physics in a gravitational field and within an accelerated
rocket: The classical Doppler effect suffices to shows that
clocks at the two ends of the rocket do not tick at the same
rate, and one can immediately deduce that lower clocks are slower
in a gravitational field. More precisely, a metric coupling
implies that in a static Newtonian potential $g_{00} = -1 + 2
U(\vec{x})/c^2+\mathcal{O}(1/c^4)$, the proper times measured by
two clocks is such that $\tau_1/\tau_2 = 1 + \left[U(\vec{x}_1)-
U(\vec{x}_2)\right]/c^2+\mathcal{O}(1/c^4)$. This has been tested
at the $2\times 10^{-4}$ level thirty years ago by flying a
hydrogen maser clock \cite{Vessot,Vessot2}, and the planned
Pharao/Aces mission \cite{pharao} should increase the precision
by two orders of magnitude.

In conclusion, the four consequences of a metric coupling have
been very well tested experimentally, notably the universality of
free fall (i.e., the relative \textit{motion of massive bodies}
in a gravitational field). Therefore, although theoretical
considerations let us expect that the Einstein equivalence
principle is violated at a fundamental level, we do know that
deviations from GR are beyond present experimental accuracy. In
the following, we will thus restrict our discussion to theories
which satisfy exactly this principle, i.e., which assume the
matter action takes the form $S_{\rm matter}[{\rm matter},
g_{\mu\nu}]$. On the other hand, we will now assume that the
kinetic term of the gravitational field, say $S_{\rm gravity}$,
is not necessarily given by the Einstein-Hilbert action of
Eq.~(\ref{eq:SGR}).

\section{Modified motion in metric theories?}
\label{sec:3}
For a given background metric $g_{\mu\nu}$, the kinetic term
$S_{\rm gravity}$ defines how gravitational waves propagate, and
the matter action $S_{\rm matter}$ how massive bodies move in
spacetime. If we assume a universal metric coupling $S_{\rm
matter}[{\rm matter}, g_{\mu\nu}]$ as in GR, we are thus tempted
to conclude that the motion of matter must be strictly the same
as in GR, and that the present lecture notes should stop here.
However, $S_{\rm gravity}$ also defines how $g_{\mu\nu}$ is
\textit{generated} by the matter distribution. Therefore, the
motion of massive bodies within this metric does actually depend
directly on the dynamics of gravity!

The clearest way to illustrate this conclusion is to integrate
away the metric tensor, i.e., to replace it in terms of its
material sources, in order to construct the so-called ``Fokker
action''. We give below a schematic derivation of its expression,
taken from \cite{DEF96}. Gauge-fixing subtleties are discussed
notably in Appendix C of \cite{DamourSchafer}. We start from an
action of the form
\begin{equation}\label{SPhiMat}
S = S_\Phi[\Phi] + S_{\rm matter}[\sigma,\Phi],
\end{equation}
where $\Phi$ denotes globally all fields participating in the
gravitational interaction, and $\sigma$ denotes the matter
sources. We also denote as $\bar\Phi[\sigma]$ a solution of the
field equation $\delta S/\delta\Phi = 0$ for given sources
$\sigma$. Let us now define the Fokker action
\begin{equation}\label{SFokker}
S_{\rm Fokker}[\sigma] \equiv
S_\Phi\left[\bar\Phi[\sigma]\right]
+ S_{\rm matter}\left[\sigma,\bar\Phi[\sigma]\right],
\end{equation}
and show that it gives the correct equations of motion for matter
$\sigma$. Indeed, its variational derivative reads
\begin{equation}
\frac{\delta S_{\rm Fokker}[\sigma] }{\delta\sigma} =
\left(\frac{\delta
S[\sigma,\Phi] }{\delta\sigma}
\right)_{\Phi = \bar\Phi[\sigma]}
+\left(\frac{\delta S[\sigma,\Phi] }{\delta\Phi}
\right)_{\Phi = \bar\Phi[\sigma]}
\frac{\delta\bar\Phi[\sigma]}{\delta\sigma},
\end{equation}
where the second term of the right-hand side vanishes because
$\bar\Phi[\sigma]$ has been chosen as a solution of $\delta
S/\delta\Phi = 0$. Therefore $\delta S_{\rm Fokker}[\sigma] /
\delta\sigma = 0$ does yield the correct equations of motion
$\delta S[\sigma,\Phi] /\delta\sigma = 0$ for matter within the
background $\Phi = \bar\Phi[\sigma]$ it consistently generates.
The most important point to notice here is that the Fokker action
(\ref{SFokker}) is \textit{not} simply given by the matter action
$S_{\rm matter}[\sigma, \Phi]$, computed in the consistent
background $\Phi = \bar\Phi[\sigma]$. Not only the
$\sigma$-dependence of this background must be taken into account
when varying the Fokker action, but its definition
(\ref{SFokker}) also depends crucially on the kinetic term (and
the nonlinear dynamics) of the field,
$S_\Phi\left[\bar\Phi[\sigma]\right]$.

\begin{figure}[b]
\sidecaption
\includegraphics[scale=.7]{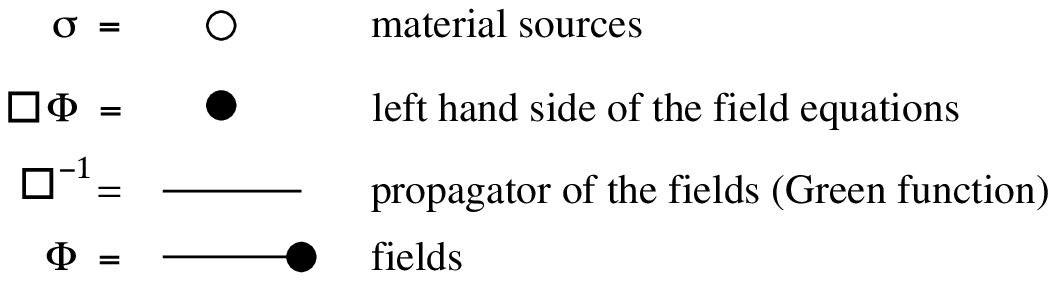}
\caption{Diagrammatic representation of matter sources, fields,
and their propagator.}
\label{fig1}
\end{figure}

\begin{figure}[t]
\includegraphics[scale=.7]{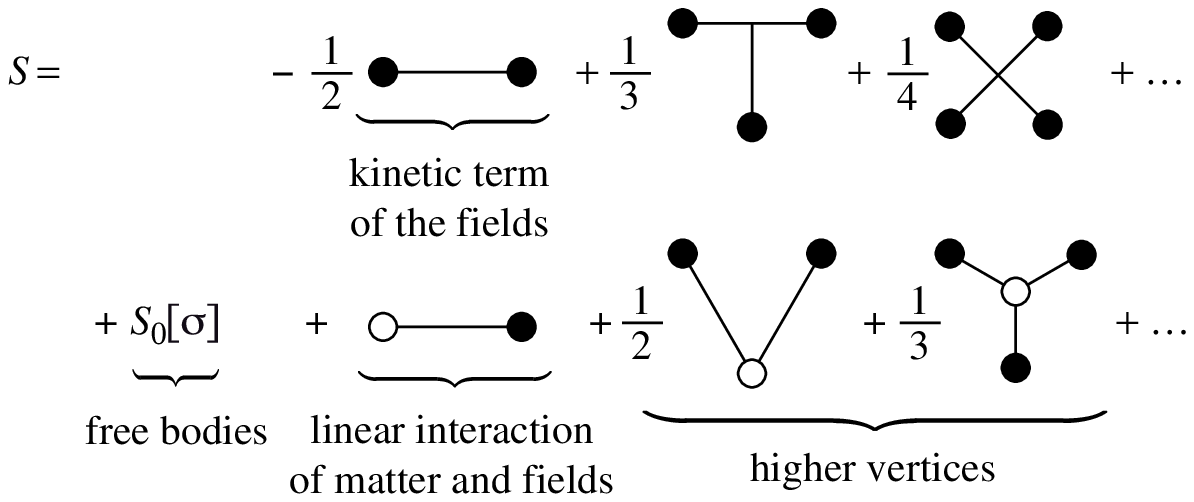}
\caption{Diagrammatic representation of the full action
(\ref{SPhiMat}) of the theory, expanded in powers of $\Phi$
(black blobs). The first line corresponds to the field action
$S_\Phi[\Phi]$, and the second one to the matter action $S_{\rm
matter}[\sigma,\Phi]$ (describing notably the matter-field
interaction).}
\label{fig2}
\end{figure}

\begin{figure}[t]
\includegraphics[scale=.6]{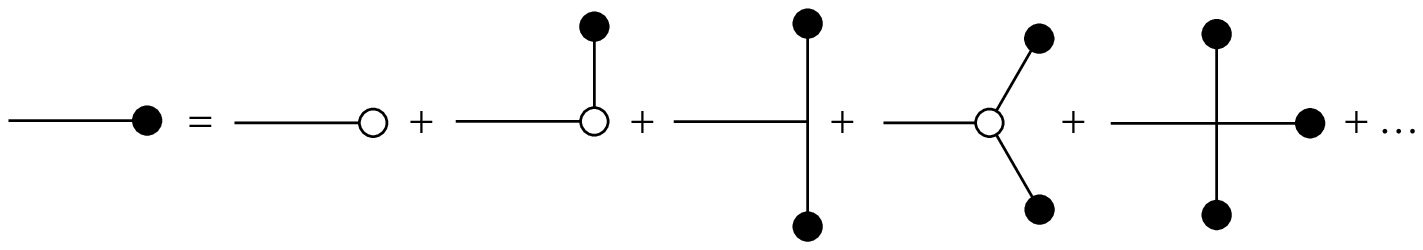}
\caption{Diagrammatic representation of the field equation
$\delta S/\delta\Phi = 0$ satisfied by $\bar\Phi[\sigma]$.}
\label{fig3}
\end{figure}

\begin{figure}[b]
\includegraphics[scale=.7]{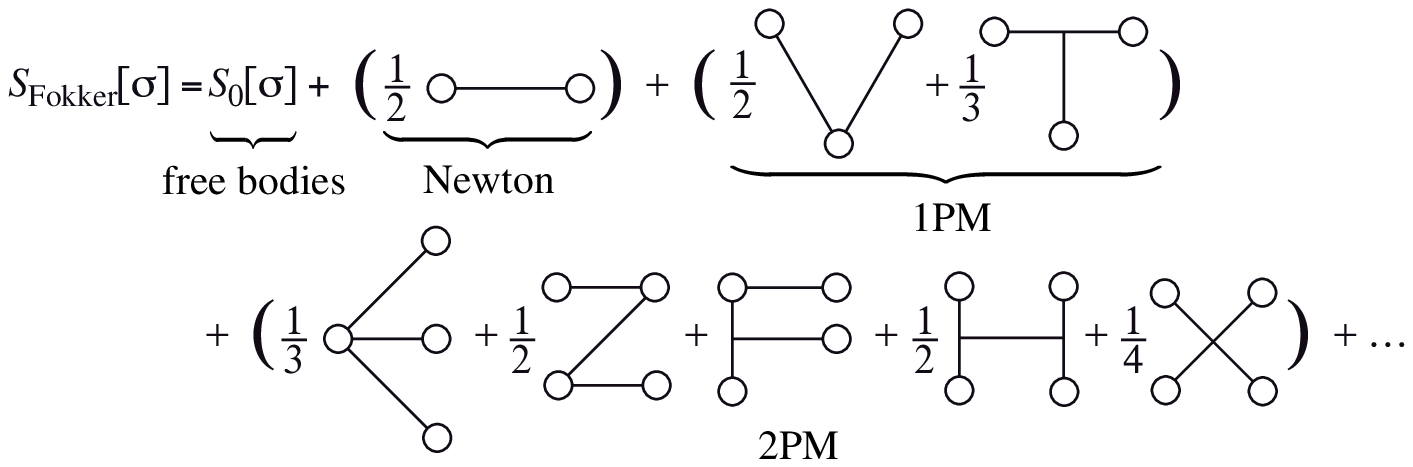}
\caption{Diagrammatic representation of the Fokker action
(\ref{SFokker}), which depends only on matter sources $\sigma$
(white blobs). The dumbbell diagram labelled ``Newton''
represents the Newtonian interaction $\propto G$, together with
all velocity-dependent relativistic corrections. The 3-blob
diagrams labelled ``1PM'' represent first post-Minkowskian
corrections, i.e., the $\mathcal{O}(G^2)$ post-Newtonian terms as
well as their full velocity dependence. The 4-blob diagrams
labelled ``2PM'' represent second post-Minkowskian corrections
$\propto G^3$.}
\label{fig4}
\end{figure}

To illustrate more vividly that the motion of massive bodies does
depend on the dynamics of the gravitational field(s), let us give
a diagrammatic representation of the above formal definition
(\ref{SFokker}) of the Fokker action. We first introduce some
symbols in Fig.~\ref{fig1}, notably white blobs for matter
sources and straight lines for field propagators.
Using this notation, the original action (\ref{SPhiMat}) may be
translated as in Fig.~\ref{fig2}, which actually \textit{defines}
the various vertices.
In this figure, the numerical factors have been chosen to
simplify the field equation satisfied by $\bar\Phi[\sigma]$,
which takes the diagrammatic form of Fig.~\ref{fig3}. This figure
tells us how to replace any black blob (field $\Phi$) by a white
blob (source $\sigma$) plus higher corrections, in which one can
again
replace iteratively black blobs by white ones plus corrections.
The Fokker action (\ref{SFokker}) is thus simply obtained by
eliminating in such a way black blobs from Fig.~\ref{fig2}, and
the result is displayed in Fig.~\ref{fig4}.
This figure clearly shows that the dynamics of the field (i.e.,
the first line of Fig.~\ref{fig2}) does contribute to that of
massive bodies. Indeed, if it had not been taken into account,
the Newtonian interaction would have been twice too large, no
``T'' diagram would have appeared in Fig.~\ref{fig4}, and the
numerical coefficients of all other diagrams would have also
changed.

For any theory of gravity, whose field dynamics is imposed by
$S_{\rm gravity}$, one may now explicitly compute the diagrams
entering Fig.~\ref{fig4}. Of course, any gauge invariance must
be fixed in order to define the field propagator as the inverse
of the quadratic kinetic term (first dumbbell diagram of
Fig.~\ref{fig2}). In the case of GR, one may for instance fix
the harmonic gauge, and the first line of Fig.~\ref{fig4}
translates\footnote{One needs to compute the integrals
represented by the various diagrams to derive expression
(\ref{EIH}). See Ref.~\cite{DEF96} for explicit diagrammatic
calculations.} as the well-known Einstein-Infeld-Hoffmann action
describing the interaction of several massive bodies labelled
$A$, $B$, \dots:
\begin{eqnarray}\label{EIH}
S_{\rm Fokker}&=&-\sum_A \int dt\, m_A c^2
\sqrt{1-{\bf v}_A^2/c^2}\nonumber\\
&&+ \frac{1}{2} \sum_{A\neq B} \int dt\,
\frac{G\, m_A m_B}{r_{AB}}\Bigl[
1+\frac{1}{2c^2} ({\bf v}_A^2+{\bf v}_B^2)
-\frac{3}{2 c^2} ({\bf
v}_A\cdot {\bf v}_B)\nonumber\\
&&\hphantom{+ \frac{1}{2} \sum_{A\neq B} \int dt\,
\frac{G\, m_A m_B}{r_{AB}}\Bigl[}
-\frac{1}{2c^2}({\bf n}_{AB}\cdot{\bf v}_A)
({\bf n}_{AB}\cdot{\bf v}_B) +\frac{\gamma^{\rm PPN}}{c^2}
({\bf v}_A -{\bf v}_B)^2 \Bigr]\nonumber\\
&&- \frac{1}{2}\sum_{B\neq A\neq C}
\int dt\, \frac{G^2 m_A m_B m_C}{r_{AB}
r_{AC}\, c^2} (2\beta^{\rm PPN}-1)
+ \mathcal{O}\left(\frac{1}{c^4}\right).
\end{eqnarray}
Here $r_{AB}$ denotes the (instantaneous) distance between bodies
$A$ and $B$, ${\bf n}_{AB}$ is the unit 3-vector pointing from
$B$ to $A$, ${\bf v}_A$ is the 3-velocity of body $A$, and a sum
over $B\neq A\neq C$ allows $B$ and $C$ to be the same body. The
first line of Eq.~(\ref{EIH}), noted $S_0[\sigma]$ in
Figs.~\ref{fig2} and \ref{fig4}, merely describes free bodies in
special relativity. The second line of Eq.~(\ref{EIH}) describes
the 2-body interaction, i.e., the dumbbell diagram of
Fig.~\ref{fig4} that we labelled ``Newton''. Its lowest-order
term is indeed the Newtonian gravitational potential, and we have
also displayed its first post-Newtonian (1PN) corrections, of
order $\mathcal{O}(v^2/c^2)$. Finally, the last line of
Eq.~(\ref{EIH}) corresponds to the ``V'' and ``T'' diagrams
labelled ``1PM'' in Fig.~\ref{fig4}, computed here at their
lowest (1PN) order.

The two coefficients $\beta^{\rm PPN}$ and $\gamma^{\rm PPN}$
entering Eq.~(\ref{EIH}) are simply equal to unity in GR. They
were introduced by Eddington \cite{Eddington} to describe
phenomenologically other possible theories of gravity, although
he did not have any specific model in mind. It happens that the
most natural alternatives to GR, scalar-tensor theories (that we
will introduce in Sec.~\ref{sec:4} below), do predict different
values for these two parameters. This comes from the fact that
massive bodies can exchange scalar particles in addition to the
usual gravitons of GR.
\begin{figure}[t]
\sidecaption[t]
\includegraphics[scale=.65]{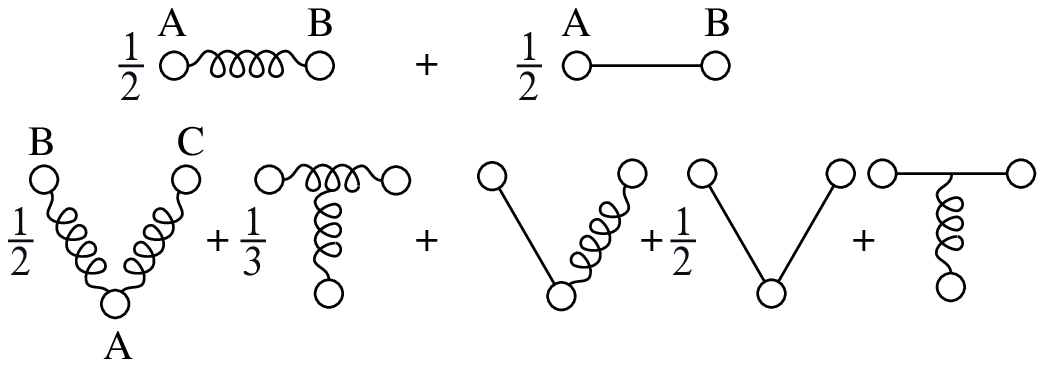}
\caption{Diagrams contributing to the $N$-body action (\ref{EIH})
at 1PN order, in scalar-tensor theories of gravity. Graviton and
scalar exchanges are represented respectively as curly and
straight lines.}
\label{fig5}
\end{figure}
If we represent gravitons as curly lines and scalar fields as
straight lines, the diagrams contributing to the Fokker action
(\ref{EIH}) are indeed displayed in Fig.~\ref{fig5}. The four
diagrams involving at least one scalar line contribute to change
the values of the Eddington parameters $\beta^{\rm PPN}$ and
$\gamma^{\rm PPN}$. We will give their explicit values in
Eq.~(\ref{PPNscal}) below, but we refer to Ref.~\cite{DEF96} for
their derivation from diagrammatic calculations.

Besides $\beta^{\rm PPN}$ and $\gamma^{\rm PPN}$, many other
parameters may actually be introduced to describe the most
general behavior of massive bodies at the first post-Newtonian
order. Under reasonable assumptions, notably that the matter
action takes the metric form $S_{\rm matter}[{\rm matter},
g_{\mu\nu}]$ and that the gravitational interaction does not
involve any specific length scale, Nordtvedt and Will
\cite{Will-Nordtvedt72} showed that 8 extra parameters are
\textit{a priori} possible, in addition to Eddington's
$\beta^{\rm PPN}$ and $\gamma^{\rm PPN}$. However, these 8
parameters vanish both in GR and in scalar-tensor gravity,
therefore we will not introduce them in these lecture notes. A
detailed presentation is available in the book \cite{Willbook}.

\section{Scalar-tensor theories of gravity}
\label{sec:4}
Among alternative theories of gravity, those which involve scalar
partners to the graviton are privileged for several reasons. Not
only their existence is predicted in all extra-dimensional
theories, but they also play a crucial role in modern cosmology
(in particular during the accelerated expansion phases of the
Universe). They are above all consistent field theories, with a
well-posed Cauchy problem, and they respect most of GR's
symmetries (notably conservation laws and local Lorentz
invariance even if a subsystem is influenced by external masses).
To simplify the discussion, we will focus on models involving a
single scalar field, although the study of tensor--multi-scalar
theories can also be done in great detail~\cite{DEF92}. We will
thus consider the class of theories defined by the
action~\cite{Bergmann,Nordtvedt70,Wagoner}
\begin{equation}
S =\frac {c^3}{16 \pi G_*}\int d^4x \sqrt{-g^*}
\left(R^*-2g_*^{\mu\nu}\partial_\mu\varphi
\partial_\nu\varphi\right) + S_{\rm matter}\left[{\rm matter} ;
g_{\mu\nu} \equiv A^2(\varphi) g^*_{\mu\nu}\right].
\label{actionST}
\end{equation}
A potential $V(\varphi)$ may also be considered in this action,
and is actually crucial in cosmology, but we will study here
solar-system-size effects and assume that the scalar-field mass
(and other self-interactions described by $V(\varphi)$) is small
enough to be negligible at this scale. The physical metric
$g_{\mu\nu}$, to which matter is universally coupled (and which
defines thus the lengths and times measured by material rods and
clocks), is the product of the Einstein metric $g^*_{\mu\nu}$
(whose kinetic term is the Einstein-Hilbert action) and a
function $A^2(\varphi)$ characterizing how matter is coupled to
the scalar field. It will be convenient to expand it around the
background value $\varphi_0$ of the scalar field far from any
massive body, as
\begin{equation}
\ln A(\varphi) = \ln A(\varphi_0)
+ \alpha_0 (\varphi-\varphi_0) +
\frac{1}{2}\beta_0(\varphi-\varphi_0)^2
+ \mathcal{O}(\varphi-\varphi_0)^3,
\label{lnA}
\end{equation}
where $\alpha_0$ defines the linear coupling constant of matter
to scalar excitations, $\beta_0$ its quadratic coupling to two
scalar lines, etc.

\subsection{Weak-field predictions}
\label{subsec:4.1}
Newtonian and post-Newtonian predictions depend only on these
first two coupling constants, $\alpha_0$ and $\beta_0$. For
instance, the effective gravitational constant between two bodies
is not given by the bare constant $G_*$ entering
action~(\ref{actionST}), but by $G = G_*(1+\alpha_0^2)$, in which
a contribution $G_*$ comes from the exchange of a (spin-2)
graviton whereas $G_* \alpha_0^2$ is due to the exchange of a
(spin-0) scalar field, each matter-scalar vertex bringing a
factor $\alpha_0$. The first line of Fig.~\ref{fig5} gives a
diagrammatic illustration of this sum. [Actually, the value of a
gravitational constant depends on the chosen units, and the
expression $G = G_* (1+\alpha_0^2)$ corresponds to the
``Einstein-frame'' representation used to write
action~(\ref{actionST}). An extra factor $A_0^2 = A(\varphi_0)^2$
enters when using the physical metric $g_{\mu\nu} = A^2(\varphi)
g^*_{\mu\nu}$ to define observable quantities, and the actual
gravitational constant which is measured reads $G_* A_0^2
(1+\alpha_0^2)$. No such extra factors $A_0$ enter the
computation of dimensionless observable quantities, like the
Eddington parameters $\beta^{\rm PPN}$ and $\gamma^{\rm PPN}$.]
\begin{figure}[b]
\sidecaption
\includegraphics[scale=.6]{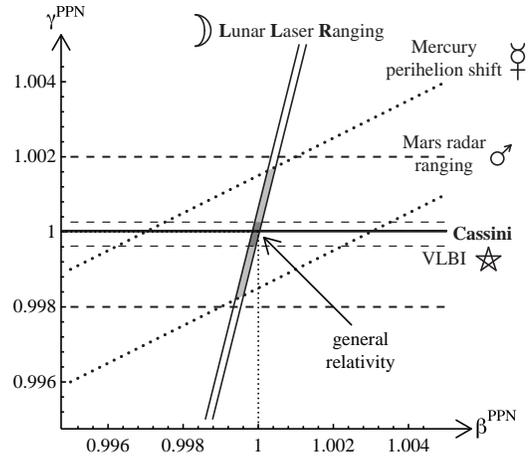}
\caption{Solar-system constraints on the post-Newtonian
parameters $\beta^{\rm PPN}$ and $\gamma^{\rm PPN}$. The allowed
region is the tiny intersection of the ``Lunar Laser Ranging''
strip with the horizontal bold line labelled ``Cassini''. General
relativity, corresponding to $\beta^{\rm PPN} = \gamma^{\rm PPN}
= 1$, is consistent with all tests.}
\label{fig6}
\end{figure}
Two kinds of post-Newtonian corrections enter the Fokker action
(\ref{EIH}): velocity-dependent terms in the 2-body interaction
(first line of Fig.~\ref{fig5}), which involve the parameter
$\gamma^{\rm PPN}$, and the lowest-order 3-body interactions
(second line of Fig.~\ref{fig5}), which involve $\beta^{\rm
PPN}$. Diagrammatic calculations \cite{DEF98} or more standard
techniques \cite{Willbook,DEF92} can be used to compute their
expressions in scalar-tensor theories:
\begin{equation}
\gamma^{\rm PPN} = 1 - \frac{2 \alpha_0^2}{1+\alpha_0^2}\,,
\qquad \beta^{\rm PPN} = 1+\frac{1}{2}\,
\frac{\alpha_0\beta_0\alpha_0}{(1+\alpha_0^2)^2}\,.
\label{PPNscal}
\end{equation}
Here again, the factor $\alpha_0^2$ comes from the exchange of a
scalar particle between two bodies, whereas
$\alpha_0\beta_0\alpha_0$ comes from a scalar exchange between
three bodies (cf.~the purely scalar ``V'' diagram of
Fig.~\ref{fig5}).

\begin{figure}[t]
\sidecaption[t]
\includegraphics[scale=.55]{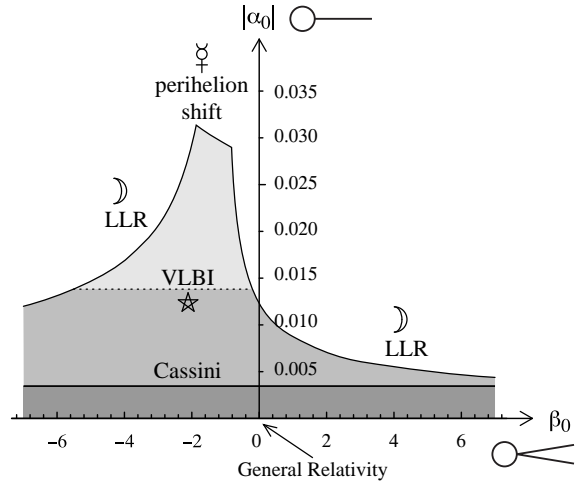}
\caption{Solar-system constraints on the matter-scalar coupling
function $\ln A(\varphi)/A_0 = \alpha_0(\varphi-\varphi_0) +
\frac{1}{2} \beta_0 (\varphi-\varphi_0)^2 +\cdots$ The allowed
region is the dark grey horizontal strip. The vertical axis
($\beta_0=0$) corresponds to Brans-Dicke theory
\cite{Brans-Dicke,Jordan,Fierz} with a parameter $2\omega_{\rm
BD}+3 = 1/\alpha_0^2$. The horizontal axis ($\alpha_0=0$)
corresponds to theories which are perturbatively equivalent to
GR, i.e., which predict strictly no deviation from it (at any
order $1/c^n$) in the weak-field conditions of the solar system.}
\label{fig7}
\end{figure}

Several solar-system observations tightly constrain these
post-Newtonian parameters to be close to 1, i.e., their general
relativistic values. The main ones are Mercury's perihelion
advance \cite{Mercury}, Lunar Laser Ranging (which allows us to
test the so-called Nordtvedt effect, i.e., whether there is a
difference between the Earth's and the Moon's accelerations
towards the Sun) \cite{Williams:2004qb}, and experiments
involving the propagation of light in the curved spacetime of the
solar system (by order of increasing accuracy: radar echo delay
between the Earth and Mars, light deflection measured by Very
Long Baseline Interferometry over the whole celestial sphere
\cite{VLBI}, and time-delay variation to the Cassini spacecraft
near solar conjunction \cite{Cassini}). These post-Newtonian
constraints are summarized in Fig.~\ref{fig6}, and the conclusion
is that GR is basically the only theory consistent with
weak-field experiments. However, when translated in terms of the
linear and quadratic coupling constants $\alpha_0$ and $\beta_0$
of matter to the scalar field, the same solar-system constraints
take the shape of Fig.~\ref{fig7}.
Therefore, the linear coupling constant $|\alpha_0|$ must be
smaller than $3\times 10^{-3}$, but we do not have any
significant constraint on $\beta_0$ [nor any higher-order vertex
entering expansion (\ref{lnA})].

\subsection{Strong-field predictions}
\label{subsec:4.2}
A qualitatively different class of constraints is obtained by
studying scalar-tensor theories in the strong-field regime, i.e.,
near compact bodies whose radius $R$ is not extremely large with
respect to their Schwarzschild radius $2Gm/c^2$. This is notably
the case when considering neutron stars, whose ratio $Gm/Rc^2
\sim 0.2$ is not far from the theoretical maximum of $0.5$ for
black holes. Although the metric is very significantly different
from the flat one inside such compact bodies and in their
immediate vicinity, their orbital velocity in a binary system may
nevertheless be small enough to perform a consistent expansion in
powers of $v^2/c^2 \sim Gm/rc^2$, where $r$ denotes the interbody
distance (as opposed to their radius $R$). This is usually called
a post-Keplerian expansion\footnote{Two different (though related)
meanings of ``post-Keplerian'' exist in the literature. We are
here considering a post-Keplerian \textit{expansion} in powers of
$v_{\rm orbital}^2/c^2$, while keeping the full nonperturbative
dependence in the gravitational self-energy $Gm/Rc^2$. On the
other hand, post-Keplerian \textit{deviations} mean relativistic
effects modifying the lowest-order Keplerian motion, like those
described in Sec.~\ref{subsec:4.3} below. Only this latter
meaning is used in GR, because its strong equivalence principle
implies that the internal structure of a body does not influence
its motion up to order $\mathcal{O}(1/c^{10})$, as recalled in
Sec.~\ref{sec:5} below.}. In such a case, one can show
\cite{Willbook,DEF92} that the predictions of scalar-tensor
theories are similar to those of weak-field conditions, with the
only difference that the matter-scalar coupling constant
$\alpha_0$ and $\beta_0$ are replaced by body-dependent
quantities, say $\alpha_A$ and $\beta_A$ for a body labelled $A$.
For instance, the effective gravitational constant describing the
lowest-order attraction between two compact bodies $A$ and $B$
reads now $G_{AB} = G_* (1+\alpha_A \alpha_B)$, instead of the
weak-field expression $G = G_* (1+\alpha_0^2)$ mentioned above.
Similarly, the post-Newtonian parameters $\beta^{\rm PPN}$ and
$\gamma^{\rm PPN}$ are replaced by body-dependent ones
$\beta^A_{BC}$ and $\gamma_{AB}$, taking the same forms as in
Eq.~(\ref{PPNscal}) but where $\alpha_0^2$ is replaced by
$\alpha_A \alpha_B$, and $\alpha_0\beta_0\alpha_0$ by
$\alpha_B\beta_A\alpha_C$ (see Ref.~\cite{DEF92} for precise
expressions). All post-Keplerian effects can thus be derived
straightforwardly, in a similar way as in the solar system. In
addition to these predictions, one may also compute the energy
loss due to the emission of gravitational waves by a binary
system. It takes the schematic form\footnote{The precise
definitions of these multipoles and their explicit expressions
may be found for instance in Sec.~6 of Ref.~\cite{DEF92}.}
\begin{eqnarray}
{{\rm Energy} \atop {\rm flux}}&=&
\left\{\frac{\rm Quadrupole}{c^5}
+\mathcal{O}\left(\frac{1}{c^7}\right)
\right\}_{\rm spin\ 2}\nonumber\\
&+&\left\{\frac{\rm Monopole}{c}
\left(0+\frac{1}{c^2}\right)^2
+\frac{\rm Dipole}{c^3}+\frac{\rm Quadrupole}{c^5}
+\mathcal{O}\left(\frac{1}{c^7}\right)
\right\}_{\rm spin\ 0}\!\!\!\!\!\!\!\!,\qquad
\label{EnergyFlux}
\end{eqnarray}
where the first line comes from the emission of usual (spin-2)
gravitons, and the second one from the emission of scalar
(spin-0) waves. Note that the dipolar term is of order
$\mathcal{O}(1/c^3)$, generically much larger than the standard
$\mathcal{O}(1/c^5)$ quadrupole of GR. As expected for a dipole,
it vanishes when considering a perfectly symmetrical binary
system, because there is no longer any privileged spatial
orientation. Its precise calculation \cite{Willbook,DEF92} shows
indeed that it involves the \textit{difference} of the scalar
charges of the two bodies, and is actually proportional to
$(\alpha_A-\alpha_B)^2$. Although the monopolar term is \textit{a
priori} of the even larger order $\mathcal{O}(1/c)$ for bodies
which are not at equilibrium (e.g. collapsing or exploding
stars), it reduces to order $\mathcal{O}(1/c^5)$ for usual
bodies, as displayed in Eq.~(\ref{EnergyFlux}), because of the
conservation of their scalar charge.

\begin{figure}[t]
\sidecaption[t]
\includegraphics[scale=.55]{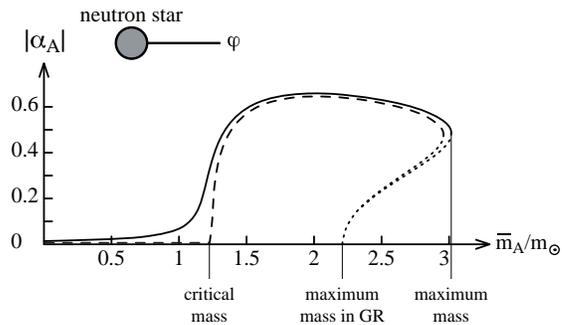}
\caption{Scalar charge $\alpha_{\rm A}$ of a neutron star versus
its baryonic mass $\overline m_{\rm A}$, for the model
$A(\varphi) = \exp(-3\varphi^2)$, i.e., $\beta_0 = -6$. The solid
line corresponds to a small value of $\alpha_0$ (namely, the VLBI
bound of Fig.~\ref{fig7}), and the dashed line to $\alpha_0 = 0$.
The dotted lines correspond to unstable configurations of the
star.}
\label{fig8}
\end{figure}

The only remaining difficulty, to derive the predictions of
scalar-tensor theories in the strong-field regime, is to compute
the body-dependent coupling constants $\alpha_A$ and $\beta_A$.
This can be done thanks to numerical integrations of the field
equations inside the bodies, as explained in
\cite{DEF93,DEF96b,DEF98}. For negative values of the parameter
$\beta_0$ entering expansion (\ref{lnA}), one shows, both
analytically and numerically, that nonperturbative effects occur
beyond a critical compactness $Gm/Rc^2$ depending on $\beta_0$.
For instance, for $\beta_0 = -6$, the linear coupling constant
$\alpha_A$ of a neutron star to the scalar field takes the values
displayed in Fig.~\ref{fig8}.
One sees that even if $\alpha_0$ is vanishingly small in the
background (and thereby in the solar system), neutron stars can
develop an order-one coupling constant to the scalar field. Their
physics and their orbital motion can thus differ significantly
from the predictions of GR, although the scalar field may have
strictly no effect in the solar system.

\subsection{Binary-pulsar tests}
\label{subsec:4.3}
Now that we know how to compute the predictions of scalar-tensor
theories even in strong-field conditions, how may we test them?
It happens that nature has provided us with fantastic objects
called pulsars. These are neutron stars (thereby very compact
objects, $Gm/Rc^2 \sim 0.2$) which are rapidly rotating and
highly magnetized, and which emit a beam of radio waves like
lighthouses. They can thus be considered as natural clocks, and
the oldest pulsars are indeed very stable ones. Therefore, a
pulsar $A$ orbiting a companion $B$ is a moving clock, the best
tool that one could dream of to test a relativistic theory.
Indeed, by precisely timing its pulse arrivals, one gets a
stroboscopic information on its orbit, and one can measure
several relativistic effects. Such effects do depend on the two
masses $m_A$, $m_B$, which are not directly measurable. However,
two different effects suffice to determine them, and a third
relativistic observable then gives a test of the theory.

For instance, in the case of the famous Hulse-Taylor binary
pulsar PSR B1913+16 \cite{1913}, three relativistic parameters
have been determined with great accuracy: (i)~the Einstein time
delay parameter $\gamma_T$, which combines the second-order
Doppler effect ($\propto v_{\rm A}^2/2 c^2$) together with the
redshift due to the companion ($\propto G m_{\rm B}/r_{\rm AB}
c^2$); (ii)~the periastron advance $\dot\omega$ ($\propto
v^2/c^2$); and (iii)~the rate of change of the orbital period,
$\dot P$, caused by gravitational radiation damping ($\propto
v^5/c^5$ in GR, but of order $v^3/c^3$ in scalar-tensor theories;
see Eq.~(\ref{EnergyFlux})). The same parameters have also been
measured for the neutron star-white dwarf binary PSR
J1141$-$6545, but with less accuracy~\cite{Bailes,Bailes2}. In
addition to these three parameters, (iv)~the ``range'' (global
factor $G m_{\rm B}/c^3$) and (v)~``shape'' (time dependence) of
the Shapiro time delay have also been determined for two other
binary pulsars, PSR B1534+12~\cite{1534} and PSR
J0737$-$3039~\cite{0737,0737b,0737c}. The latter system is
particularly interesting because both bodies have been detected
as pulsars. Since their independent timing gives us the
(projected) size of their respective orbits, the ratio of these
sizes provides a direct measure of (vi)~the mass ratio $m_{\rm
A}/m_{\rm B}\approx 1.07$. In other words, 6 relativistic
parameters have been measured for the double pulsar PSR
J0737$-$3039. After using two of them to determine the masses
$m_{\rm A}$ and $m_{\rm B}$, this system thereby provides $6-2 =
4$ tests of relativistic gravity in strong-field conditions.

\begin{figure}[b]
\includegraphics[scale=.67]{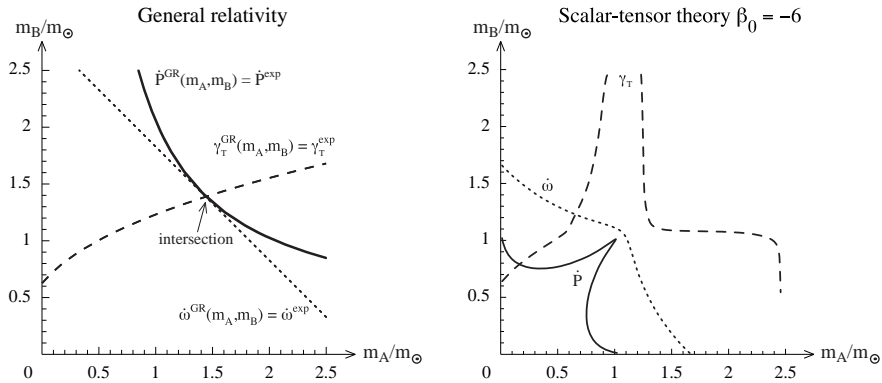}
\caption{Mass plane ($m_A=$ pulsar, $m_B =$ companion) of the
Hulse-Taylor binary pulsar PSR B1913+16 in general relativity
(left panel) and for a scalar-tensor theory with $\beta_0 = -6$
(right panel). The widths of the lines are larger than $1\sigma$
error bars. While GR passes the test with flying colors, the
value $\beta_0 = -6$ is ruled out. [Actually, the right panel
is plotted for a specific small value of $|\alpha_0| \approx
10^{-2}$, but the three curves keep similar shapes whatever
$\alpha_0$, even vanishingly small, and they never have any
common intersection for $\beta_0 = -6$.]}
\label{fig9}
\end{figure}

\begin{figure}[t]
\sidecaption[t]
\includegraphics[scale=.55]{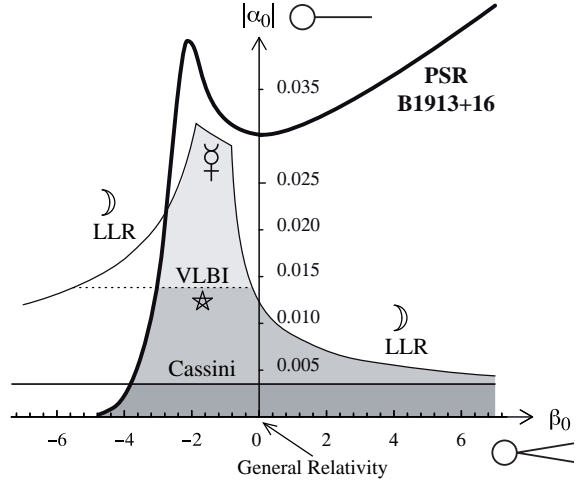}
\caption{Same theory plane $(|\alpha_0|,\beta_0)$ as in
Fig.~\ref{fig7}, but taking now into account the constraints
imposed by the Hulse-Taylor binary pulsar (bold line). LLR stands
as before for ``lunar laser ranging'' and VLBI for ``very long
baseline interferometry''. The allowed region is the dark grey
one. While solar system tests impose a small value of
$|\alpha_0|$, binary pulsars impose the orthogonal constraint
$\beta_0 > -4.5$.}
\label{fig10}
\end{figure}

The clearest way to illustrate these tests is to plot the various
experimental constraints in the mass plane $(m_{\rm A},m_{\rm
B})$, for a given theory of gravity. Any theory indeed predicts
the expressions of the various timing parameters in terms of
these unknown masses and other Keplerian observables, such
as the orbital period and the eccentricity. The equations
$\textit{predictions}(m_{\rm A},m_{\rm B}) =\textit{observed
values}$ thereby define different curves in the mass plane, or
rather different \textit{strips} if one takes into account
experimental errors. If these strips have a common intersection,
there exists a pair of masses which is consistent with all
observables, and the theory is confirmed. On the other hand, if
the strips do not meet simultaneously, the theory is ruled out.
Figure~\ref{fig9} displays this mass plane for the Hulse-Taylor
binary pulsar.
Its left panel shows that GR is superbly consistent with these
data. Its right panel illustrates that the three strips can be
significantly deformed in scalar-tensor theories, because scalar
exchanges between the pulsar and its companion modify all
theoretical predictions. In the displayed case, corresponding to
a quadratic matter-scalar coupling constant $\beta_0 = -6$ (as in
Fig.~\ref{fig8}), the strips do not meet simultaneously and the
theory is thus excluded. On the contrary, they may have a
common intersection in other scalar-tensor theories, even if it
does not correspond to the same values of the masses $m_A$ and
$m_B$ that were consistent with GR. The allowed region of the
theory space $(|\alpha_0|, \beta_0)$ is displayed in
Fig.~\ref{fig10}.

\begin{figure}[ht]
\sidecaption[t]
\includegraphics[scale=.48]{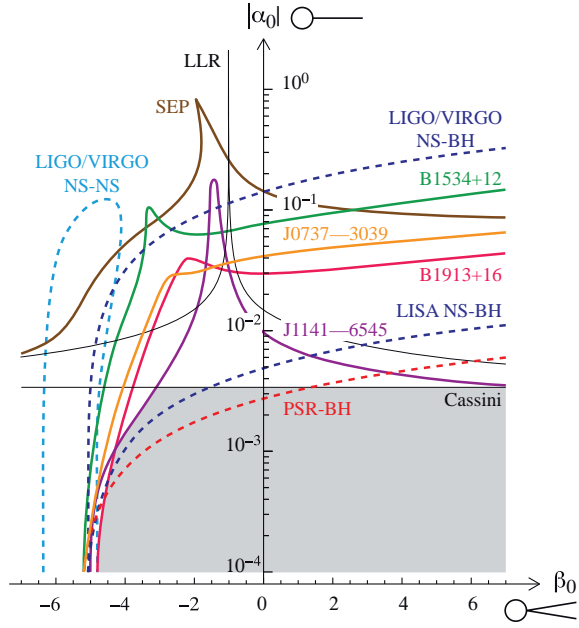}
\caption{Same theory plane as in Fig.~\ref{fig10}, but now with a
logarithmic scale for the linear matter-scalar coupling constant
$|\alpha_0|$. This plot displays solar-system and several
binary-pulsar constraints, using the latest published data. The
allowed region is shaded. The curve labeled SEP corresponds to
tests of the ``strong equivalence principle'' using a set of
neutron star-white dwarf low-eccentricity binaries. The dashed
lines corresponds to expected constraints if we find a
relativistic pulsar-black hole (PSR-BH) binary, or when
gravity-wave antennas detect the coalescence of double-neutron
star (NS-NS) or neutron star-black hole (NS-BH) binaries.}
\label{fig11}
\end{figure}
As mentioned above, several other relativistic binary pulsars are
presently known, and Fig.~\ref{fig11} displays their simultaneous
constraints on this theory plane
\cite{DEF96b,DEF98,MGX,Damour:2007uf,DEF09}. To clarify this
plot, we have used a logarithmic scale for the vertical
($|\alpha_0|$) axis. The drawback is that GR, corresponding to
$\alpha_0 = \beta_0 = 0$, is sent down to infinity, but the
important point to recall is that it does lie within the allowed
(grey) region. Figures~\ref{fig10} and \ref{fig11} illustrate
vividly the \textit{qualitative} difference between solar-system
and binary-pulsar observations. Indeed, while weak-field tests
only constrain the linear coupling constant $|\alpha_0|$ to be
small without giving much information about the quadratic
coupling constant $\beta_0$, binary pulsars impose $\beta_0 >
-4.5$ even for a vanishingly small $\alpha_0$. This constraint is
due to the spontaneous scalarization of neutron stars which
occurs when $-\beta_0$ is large enough, and which was illustrated
in Fig.~\ref{fig8} above. Equations~(\ref{PPNscal}) allow us to
rewrite this inequality in terms of the Eddington parameters
$\beta^{\rm PPN}$ and $\gamma^{\rm PPN}$, which are both
consistent with 1 in the solar system. One finds \begin{equation}
\frac{\beta^{\rm PPN}-1}{\gamma^{\rm PPN}-1} < 1.1\,.
\end{equation} The singular ($0/0$) nature of this ratio
underlines why such a conclusion could not be obtained in
weak-field experiments.

\subsection{Black holes in scalar-tensor gravity}
\label{subsec:4.4}
In Fig.~\ref{fig11} is displayed the order of magnitude of the
tight constraint that can be expected if we are lucky enough to
detect a relativistic pulsar-black hole binary. However, let us
comment on the behavior of black holes in scalar-tensor theories.
Since we saw in Fig.~\ref{fig8} above that nonperturbative
effects can occur in strong-field conditions, one should
\textit{a priori} expect even larger deviations from GR for black
holes (extreme compactness $Gm/Rc^2 = 0.5$) than for neutron
stars (large compactness $Gm/Rc^2 \sim 0.2$). However the
so-called no-hair theorem
\cite{nohair1,nohair2,nohair3,nohair4,nohair5,DEF92} shows that
black holes must have a strictly vanishing scalar charge,
$\alpha_{BH} = 0$. The basic idea is that otherwise the scalar
field $\varphi$ would diverge at the horizon, and this would be
an unphysical solution. A first consequence is that a collapsing
star must radiate away its scalar charge when forming a black
hole. This is related to the generically large $\mathcal{O}(1/c)$
monopolar radiation of scalar waves predicted for non-equilibrium
configurations, as discussed below Eq.~(\ref{EnergyFlux}). But
the second crucial consequence is that black holes, once formed
and stabilized, are not coupled at all to the scalar field, and
therefore behave exactly as in GR: They generate the same
solution for the Einstein metric $g^*_{\mu\nu}$, do not excite
the scalar field $\varphi$, and move within the curved geometry
of $g^*_{\mu\nu}$ as in GR. The conclusion is therefore that
there is strictly no observable scalar-field effect in a binary
black-hole system.

Of course, there do exist significant perturbations caused by the
scalar field during the short time of the black-hole formation or
when it captures a star (see e.g. \cite{Chauvineau:2004va}),
because of the emission of a generically large amount of energy
via scalar waves. Similarly, if one assumes that there exists a
non-constant background of scalar field $\varphi_0(x)$, then its
own energy momentum tensor $T_{\mu\nu}$ contributes to the
curvature of the Einstein metric $g^*_{\mu\nu}$, and even black
holes would thus indirectly feel its presence. However, this
would not be a consequence of the modification of gravity itself,
but of the assumption of a non-trivial background $T_{\mu\nu}$.
Even within pure GR, one could also have assumed that black holes
move within a non-trivial background, caused for instance by the
presence of dark matter or some large gravitational waves, and
one would predict then a different motion than in a trivial
background. One should thus qualify our conclusion above: Given
some precise boundary conditions, for instance asymptotic
flatness and no incoming radiation, black holes at equilibrium
behave exactly as in GR.

It remains now to understand how a pulsar-black hole binary can
allow us to constrain scalar-tensor theories as in
Fig.~\ref{fig11}. The reason is that one of the two bodies is
\textit{not} a black hole, but a neutron star with scalar charge
$\alpha_A\neq 0$. Let us even recall that massive enough neutron
stars can develop order-one scalar charges as in Fig.~\ref{fig8},
even if matter did not feel at all the scalar field in the solar
system ($\alpha_0 = 0$). A pulsar-black hole binary must
therefore emit a large amount of dipolar waves $\propto
(\alpha_A-\alpha_{BH})^2/c^3 = \alpha_A^2/c^3 =
\mathcal{O}(1/c^3)$, as given by Eq.~(\ref{EnergyFlux}), and this
can be several orders of magnitude larger than the usual
$\mathcal{O}(1/c^5)$ quadrupolar radiation predicted by GR. Any
pulsar-black hole binary whose variation of the orbital period,
$\dot P$, is consistent with GR will thus tightly constrain
scalar-tensor models.

\section{Extended bodies}
\label{sec:5}
In this section, we will discuss finite-size effects on the
motion of massive bodies, in GR and in scalar-tensor theories
\cite{Nordtvedt:1994,DEF98,Goldberger:2004jt}. It is still
convenient to describe the position of such extended bodies by
using one point in their interior, for instance their approximate
center of mass. In other words, we will skeletonize the extended
body's worldtube as a unique worldline, say $x_c$. However, the
action (\ref{eq:Spp}) describing the motion of a point particle
cannot remain valid to all orders, because the metric
$g_{\mu\nu}(x_{\rm left})$ on one side of the body is not
strictly the same as $g_{\mu\nu}(x_{\rm right})$ on the other
side. By expanding the metric around its value at $x_c$, it is
thus clear that an effective action describing the motion of an
extended body must depend on \textit{derivatives} of the metric,
and since we wish to construct a covariant expression, such
derivatives must be built from contractions of the curvature
tensor, its covariant derivatives, or any product of them.
Moreover, if the body is nonspinning and spherical when
isolated, it does not have any privileged direction, and the only
4-vector available to contract possible free indices is the
4-velocity $u^\mu= dx^\mu/ds$ of the point $x_c$. The first
couplings to curvature that one may think of are thus of the form
\cite{Goldberger:2004jt}
\begin{equation}
\label{eq:12}
S_{\rm extended\ body} = S_{\rm point\ particle} +
\int \left(k_1 R + k_2 R_{\mu\nu} u^\mu u^\nu
+ \cdots\right) c\, ds,
\end{equation}
where $k_1$ and $k_2$ denote body-dependent form factors (no term
$R_{\mu\nu\rho\sigma} u^\mu u^\nu u^\rho u^\sigma$ is written
since it vanishes because of the symmetry properties of the
Riemann tensor). However, as recalled in \cite{DEF98,BDEI05}, any
perturbative contribution to an action, say $S[\psi]$, which is
proportional to its lowest-order field equations, $\delta
S/\delta \psi$, is equivalent to a local field redefinition.
Indeed, $S[\psi+\varepsilon] = S[\psi] + \varepsilon \delta
S/\delta \psi +\mathcal{O}(\varepsilon^2)$, where the small
quantity $\varepsilon$ may be itself a functional of the field
$\psi$. Therefore, up to \textit{local} redefinitions of the
worldline $x_c$ and the metric $g_{\mu\nu}$ inside the body,
which do not change the observable effects encoded in the metric
$g_{\mu\nu}(x_{\rm obs})$ at the observer's location, one may
replace $R$ and $R_{\mu\nu}$ in action~(\ref{eq:12}) above by
their sources. Outside the extended body, such couplings to $R$
or $R_{\mu\nu}$ have thus strictly no observable effect (see
\cite{BDEI05} for a discussion of the field redefinitions inside
the body even when it is skeletonized as a point particle).

The first finite-size observable effects must therefore involve
the Weyl tensor $C_{\mu\nu\rho\sigma}$, which does not vanish
outside the body. Because of the symmetry properties of this
tensor, the lowest-order terms one may construct from it take the
form \cite{Goldberger:2004jt}
\begin{eqnarray}
\label{eq:13}
S_{\rm extended\ body} &=& S_{\rm point\ particle}+
\int \Bigl(k_3\, C_{\mu\nu\rho\sigma}^2
+ k_4\, C_{\mu\nu\rho\alpha}\,
C^{\mu\nu\rho}_{\phantom{\mu\nu\rho}\beta}
\,u^\alpha u^\beta\nonumber\\
&& + k_5\, C_{\mu\alpha\nu\beta}\,
C^{\mu\phantom{\gamma}\nu}_{\phantom{\mu}\gamma
\phantom{\nu}\delta}
\,u^\alpha u^\beta u^\gamma u^\delta
+ \cdots\Bigr) c\, ds.
\end{eqnarray}
By comparing the dimensions of $S_{\rm point\ particle} = -\int
mc\, ds$ and these couplings to the Weyl tensor, we expect $k\,
C^2 \sim m$ and therefore $k \sim m\, R_E^4$ (where $k$ denotes
any of the form factors $k_3$, $k_4$ or $k_5$, and $R_E$ means
the radius of the Extended body, here written with an index to
avoid a confusion with the curvature scalar). In conclusion, for
a compact body such that $Gm/R_E c^2 \sim 1$, we expect the form
factors $k$ to be of order $m (G m /c^2)^4$, and the corrections
to its motion will be proportional to $k\, c^2 C^2 = k\,
c^2\mathcal{O}(1/c^4) = \mathcal{O}(1/c^{10})$. Therefore, finite
size effects start at the fifth post-Newtonian (5PN) order,
consistently with the Newtonian reasoning. Indeed, if one
considers two bodies $A$ and $B$ which are spherical when
isolated, one can show that $B$ is deformed as an ellipsoid by
the tidal forces caused by $A$, and this deformation induces an
extra force $\sim (G m_A/r_{AB}^2)(R_B/r_{AB})^5$ felt by $A$. If
$B$ is a compact body, i.e., $G m_B / R_B c^2 \sim 1$, one
concludes that the extra force felt by body $A$ is of order
$\mathcal{O}(1/c^{10})$. We recovered above the same conclusion
within GR thanks to a very simple dimensional argument. Note
however that the explicit calculation is more involved than our
estimate of the order of magnitude, because one must take into
account the Weyl tensor generated by the \textit{two} (or more)
bodies of the system.

Let us now follow the same dimensional reasoning within
scalar-tensor theories of gravity. We assumed in
Eq.~(\ref{actionST}) that matter is coupled to the physical
metric $g_{\mu\nu} = A^2(\varphi) g^*_{\mu\nu}$, therefore the
action of a point particle reads
\begin{equation}
\label{eq:14}
S_{\rm point\ particle} =
-\int mc \sqrt{-g_{\mu\nu} dx^\mu dx^\nu}
=-\int A(\varphi) m c \sqrt{-g^*_{\mu\nu} dx^\mu dx^\nu}.
\end{equation}
When studying the motion of such a point particle in the Einstein
metric $g^*_{\mu\nu}$, it behaves thus as if it had a
scalar-dependent mass $m_*(\varphi) \equiv A(\varphi) m$.
[Incidentally, this provides us with another definition of the
scalar charge, $\alpha = d \ln m_*(\varphi)/d\varphi$.] If we
consider now an extended body in scalar-tensor gravity, the
function $m_*(\varphi)$ must be replaced by a
\textit{functional}, i.e., it can depend on the (multiple)
derivatives of both the metric and the scalar field. Let us
restrict again our discussion to nonspinning bodies which are
spherical when isolated. Then the derivative expansion of this
mass functional can be written as \cite{DEF98}
\begin{eqnarray}
\label{eq:15}
{m[\varphi,g^*_{\mu\nu}]} &=&
m(\varphi)
+ I(\varphi) R^*
+ J(\varphi) R^*_{\mu\nu}\, u_*^\mu u_*^\nu
+ K(\varphi) \Box^* \varphi
+ L(\varphi) \nabla^*_\mu\partial_\nu\varphi\,
u_*^\mu u_*^\nu\nonumber\\
&&+ M(\varphi) \partial_\mu\varphi\partial_\nu\varphi\,
u_*^\mu u_*^\nu
+ N(\varphi) g_*^{\mu\nu} \partial_\mu\varphi
\partial_\nu\varphi
+\cdots,
\end{eqnarray}
where star-indices mean that we are using the Einstein metric
$g^*_{\mu\nu}$ to compute curvature tensors, covariant
derivatives and the unit velocity $u_*^\mu \equiv dx^\mu/ds^*$,
and where $I(\varphi)$, \dots, $N(\varphi)$ are body-dependent
form factors encoding how the extended body feels the various
second derivatives of the fields $\varphi$ and $g^*_{\mu\nu}$.
Fortunately, by using as in GR the lowest-order field equations
together with local field and worldline redefinitions, this
long expression reduces to a mere
\begin{equation}
\label{eq:16}
{m[\varphi,g^*_{\mu\nu}]} =
m(\varphi)
+ N_{\rm new}(\varphi) g_*^{\mu\nu}
\partial_\mu\varphi\partial_\nu\varphi
+ \hbox{higher post-Keplerian orders},
\end{equation}
where $N_{\rm new}(\varphi)$ is a linear combination of the
previous $N(\varphi)$, $L(\varphi)$ and $I(\varphi)$. We already
notice that the lowest-order finite-size effects involve only two
derivatives of the scalar field, whereas action~(\ref{eq:13}) in
GR involved the square of a \textit{second} derivative of the
metric. As above, let us invoke dimensional analysis to
deduce that $N_{\rm new} \sim m R_E^2$, where $R_E$ still denotes
the radius of the extended body. This is consistent with the
actual calculation performed in \cite{Nordtvedt:1994} for weakly
self-gravitating bodies, where it was proven that $N_{\rm new} =
\frac{1}{6}\beta_0 \times (\hbox{Inertia moment})$. For compact
(therefore strongly self-gravitating) bodies, such that $Gm/R_E
c^2 \sim 1$, one thus expects observable effects of order $Nc^2
(\partial \varphi)^2 = \mathcal{O}(1/c^2) R_E^2 =
\mathcal{O}(1/c^6)$, i.e., of the third post-Newtonian (3PN)
order. Finite-size effects are thus much larger in
scalar-tensor theories\footnote{These larger finite-size
effects are due to the fact that a spin-0 scalar field can
couple to the spherical inertia moment of a body, contrary to a
spin-2 graviton. They should not be confused with the
violation of the strong equivalence principle, which
\textit{also} occurs in scalar-tensor theories because
all form factors $m(\varphi)$, $N(\varphi)$, \dots\ depend
on the body's self-energy. Regardless of its finite size,
the motion of a self-gravitating body in a uniform exterior
gravitational field depends thus on its internal structure.}
than in GR, where they were of the 5PN order,
$\mathcal{O}(1/c^{10})$. Moreover, when nonperturbative
strong-field effects develop as in Fig.~\ref{fig8} above, one
actually expects that $N_{\rm new} \sim \beta_E \times
(\hbox{Inertia moment})$ can be of order unity, because the
coupling constant $\beta_E$ of the body to two scalar lines can
be extremely large with respect to the bare $\beta_0$.
Therefore, finite-size effects in scalar-tensor gravity should
actually be considered as \textit{first} post-Keplerian,
$\mathcal{O}(v_{\rm orbital}^2/c^2)$, perturbations of the
motion, as compared to the fifth post-Newtonian order in GR.

\section{Modified Newtonian dynamics}
\label{sec:6}
The existence of dark matter (a pressureless and non-interacting
fluid detected only by its gravitational influence) is suggested
by several observations. For instance, type-Ia supernova data
\cite{Tonry:2003zg,Knop:2003iy,Riess:2004nr,Astier:2005qq} are
consistent with a present acceleration of the expansion of the
Universe, and tell us that the dark energy density should be of
order $\Omega_\Lambda \approx 0.7$. On the other hand, the
position of the first acoustic peak of the cosmic microwave
background spectrum \cite{Spergel:2003cb,Spergel:2006hy} is
consistent with a spatially flat Universe, i.e., $\Omega_\Lambda
+ \Omega_{\rm matter} \approx 1$. Combining these two pieces of
information, one thus deduce that the matter density should be
$\Omega_{\rm matter} \approx 0.3$, a value at least one order of
magnitude larger than all our estimates of baryonic matter in the
Universe (for instance $\Omega_{\rm baryons} \approx 0.04$
derived from Big-Bang nucleosynthesis). Therefore, most of the
cosmological matter should be non-baryonic, i.e., ``dark''
because it does not interact significantly with photons. Another,
independent, evidence for the existence of dark matter is the
flat rotation curves of clusters and galaxies~\cite{Rubin:1978}:
The velocities of outer stars tend toward a constant value
(depending on the galaxy or cluster), instead of going
asymptotically to zero as expected in Newtonian theory (recall
that Neptune is much slower than Mercury in the solar system). If
Newton's law is assumed to be valid, such nonvanishing asymptotic
velocities imply the existence of much more matter than within
the stars and the gas. Independently of these experimental
evidences, we also have many theoretical candidates for dark
matter, notably the class of neutralinos occurring in
supersymmetric theories (see e.g. \cite{Boehm:2003ha}), and
numerical simulations of structure formation have obtained great
successes while incorporating dark matter (see e.g.
\cite{Hayashi:2004}).

However, this unknown fluid might actually be an artifact of our
interpretation of experimental data with a Newtonian viewpoint.
It is thus worth examining whether the gravitational $1/r^2$ law
could be modified at large distances, instead of invoking the
existence of dark matter. In 1983, Milgrom realized that galaxy
rotation curves could be fitted with a very simple recipe, that
he called Modified Newtonian Dynamics (MOND)
\cite{Milgrom:1983ca}. It does not involve any mass scale nor
distance scale, but an \textit{acceleration} scale denoted as
$a_0$ [not to be confused with the linear matter-scalar coupling
constant $\alpha_0$, defined in Eq.~(\ref{lnA}) above]. Milgrom
assumed that the acceleration $a$ of a test particle caused by a
mass $M$ should read
\begin{equation}
\begin{array}{cccccl}
a & = & a_N & = & \displaystyle \frac{GM}{r^2} &
\quad{\rm if }~a>a_0,\\
&&&&&\\
a & = & \sqrt{a_0 a_N} & = &
\displaystyle \frac{\sqrt{GMa_0}}{r}
&\quad{\rm if }~a<a_0,
\end{array}
\label{MOND}
\end{equation}
where $a_N$ denotes the usual Newtonian expression. This
phenomenological law happens to fit remarkably well galaxy
rotation curves \cite{Sanders:2002pf}, for a universal constant
$a_0 \approx 1.2\times 10^{-10}\,{\rm m.s}^{-2}$. [However,
galaxy clusters require either another value of this constant, or
some amount of dark matter, for instance in the form of massive
neutrinos.] Moreover, it automatically recovers the Tully-Fisher
law~\cite{Tully:1977fu} $v_\infty^4 \propto M$, where $M$ denotes
the baryonic mass of a galaxy, and $v_\infty$ the asymptotic
velocity of visible matter in its outer region. The MOND
assumption (\ref{MOND}) would also explain in an obvious way why
dark matter profiles seem to be tightly correlated to the
baryonic ones \cite{McGaugh:2005er}.

However, reproducing the simple law (\ref{MOND}) in a consistent
relativistic field theory happens to be quite difficult. As
mentioned in Sec.~\ref{sec:2} above, Milgrom explored ``modified
inertia'' models \cite{Milgrom:1992hr,Milgrom:1998sy}, in which
the action of a point particle is assumed to depend nonlocally on
all the time derivatives $d^n {\bf x}/dt^n$ of its position ${\bf
x}$. In these lecture notes, we focus on local field theories
which ``modify gravity'', i.e., which assume that the kinetic
term of the metric $g_{\mu\nu}$ (to which matter is universally
coupled) is not the standard Einstein-Hilbert action
(\ref{eq:SGR}).

\subsection{Mass-dependent models?}
\label{subsec:6.1}
It is actually very easy to devise a model which predicts a force
$\propto 1/r$. Indeed, let us consider a mere scalar field
$\varphi$ in flat spacetime, with a potential $V(\varphi) = -2
a^2 e^{-b\varphi}$, where $a$ and $b$ are two constants. In a
static and spherically symmetric situation, its field equation
$\Delta \varphi = V'(\varphi)$ then gives the obvious solution
$\varphi = (2/b) \ln (abr)$. If we now assume that matter is
linearly coupled to $\varphi$, it will feel a force $\propto
\partial_i \varphi$, i.e., the $1/r$ law we are looking for.
However, this simple model presents two very serious problems.
The first one is that the potential $V(\varphi) = -2 a^2
e^{-b\varphi}$ is unbounded from below, and therefore that the
theory is unstable. Actually, we will see that stability is
indeed a generic difficulty of all models trying to reproduce the
MOND dynamics. The second problem of this naive model is that it
predicts a \textit{constant} coefficient $2/b$ for the solution
of the scalar field, instead of the factor $\sqrt{M}$ entering
the second line of (\ref{MOND}). It happens that many models
proposed in the literature do behave in the same way, although
their more complicated writings hide the problem\footnote{See
Sec.~II.B of Ref.~\cite{BEF07} for a critical discussion of
various such mass-dependent models.}. The trick used by the
corresponding authors is merely to set $b \propto 1/\sqrt{M}$ in
the action of the theory. In other words, they are considering a
\textit{different} theory for each galaxy $M$! It should be noted
that the mass of an object is not a local quantity: It is the
integral of the matter density over a particular region, whose
symmetry plays also an important role. One of the most difficult
steps in building a consistent theory of MOND is thus precisely
to be able to predict this factor $\sqrt{M}$. Moreover, if one
defined a potential $V(\varphi)$ as above, in terms of an
integral of the matter density giving access to $M$, then it
would mean that matter is coupled to $\varphi$ in a highly
nonlinear and nonlocal way, therefore the force it would feel
would be much more complicated than the naive gradient
$\partial_i \varphi$ assumed above. In the following, we will
consider local field theories whose actions do not depend on $M$,
but only on the constants $G$, $c$ and $a_0$.

\subsection{Aquadratic Lagrangians or k-essence}
\label{subsec:6.2}
One of the most promising frameworks to reproduce the MOND
dynamics is a generalization of the scalar-tensor theories we
considered in Sec.~\ref{sec:4} above. Their action takes the form
\cite{Bekenstein:1992pj,Bekenstein:1993fs,Sanders:1996wk,
Bekenstein:2004ne,Bekenstein:2004ca,Sanders:2005vd}
\begin{eqnarray}
S &=& \frac{c^3}{16\pi G_*}\int d^4 x\sqrt{-g^*} \left\{
R^* - 2\, f\left(g_*^{\mu\nu}\partial_\mu\varphi
\partial_\nu\varphi\right)\right\}
\nonumber\\
&&+ S_{\rm matter}\left[{\rm matter}; g_{\mu\nu}
\equiv~A^2(\varphi) g^*_{\mu\nu}
+ B(\varphi) U_\mu U_\nu\right].
\label{TeVeS}
\end{eqnarray}
The first crucial difference with action (\ref{actionST}) is that
the kinetic term of the scalar field is now a \textit{function}
of the standard quadratic term $(\partial_\mu \varphi)^2$.
Reference \cite{Bekenstein:1984tv} showed that this suffices to
reproduce the MOND law (\ref{MOND}), including the important
$\sqrt{M}$ coefficient, provided $f(x) \propto x^{3/2}$ for small
values of $x$ (MOND regime) and $f(x) \rightarrow x$ for large
$x$ (Newtonian regime). Such aquadratic kinetic terms have also
been analyzed later in the cosmological context, under the names
of k-inflation \cite{Armendariz-Picon:1999rj} or k-essence
\cite{Chiba:1999ka,Armendariz-Picon:2000ah} (the letter k meaning
that their dynamics is kinetic dominated).

As in action (\ref{actionST}), matter is assumed to be coupled to
the scalar field via the function $A^2(\varphi)$ entering the
definition of the physical metric $g_{\mu\nu}$. This ensures that
test particles will undergo an extra acceleration caused by the
scalar field. But the second difference with (\ref{actionST}) is
the presence of the non-conformal term $B(\varphi) U_\mu U_\nu$
entering the definition of $g_{\mu\nu}$. Here $U_\mu$ is a vector
field, which can have either its own kinetic term, or can be
simply chosen as $U_\mu = \partial_\mu \varphi$. This term is
necessary to reproduce the light deflection caused by galaxies or
clusters, which happens to be consistent with the prediction of
GR in presence of a dark matter halo
\cite{Fort:1994,Mellier:1998pk,Bartelmann:1999yn}. Indeed, light
is totally insensitive to the conformal factor $A^2(\varphi)$
relating the physical and Einstein metrics in action
(\ref{actionST}). The simplest way to understand it is to recall
that a photon's null geodesic satisfies $ds^2 = 0 \Leftrightarrow
A^2(\varphi) ds_*^2 = 0 \Leftrightarrow ds_*^2 = 0$, therefore
the photon propagates in the Einstein metric $g^*_{\mu\nu}$
without feeling the presence of the scalar field. [To be more
precise, the photon does feel indirectly the presence of the
scalar field via the extra curvature of $g^*_{\mu\nu}$ caused by
its energy-momentum tensor, but this is a higher post-Newtonian
effect.] The only way to impose that the MOND potential (encoded
in the scalar field $\varphi$) deflect light is thus to relate
the physical metric $g_{\mu\nu}$ to the Einstein one in a
non-conformal way, as in (\ref{TeVeS}). This idea dates back to
Ni's ``stratified'' theory of gravity \cite{Ni,Willbook}.

Obviously, several conditions must be imposed on the functions
entering action (\ref{TeVeS}) to warrant that the theory is
stable and that it has a well-posed Cauchy problem. For instance,
it is clear that $f(x) = x$ defines a standard kinetic term for
the scalar field, whereas $f(x) = -x$ would define a
negative-energy (ghost) mode. In order for the Hamiltonian to be
bounded by below and for the scalar-field equations to be
hyperbolic, one can actually show that the function $f$ must
satisfy the two conditions \cite{Aharonov:1969vu,BEF07}
\begin{equation}
\forall x,\quad f'(x) > 0,\qquad {\rm and}
\qquad \forall x,\quad 2\,x\, f''(x) + f'(x) > 0.
\label{conditions}
\end{equation}
One also notices that gravitons are faster than scalar particles
when $f''(x) < 0$, and slower when $f''(x) > 0$. It has been
argued in \cite{Adams:2006sv} that gravitons should be the
fastest modes, otherwise causal paradoxes can be constructed.
However, Refs.~\cite{Bruneton:2006gf,Babichev:2007dw,BEF07}
concluded that conditions (\ref{conditions}) suffice for
causality to be preserved, because they ensure that the widest
causal cone always remains a cone, and never opens totally.

The analogues of conditions (\ref{conditions}) become much more
complicated within matter, where the two other functions
$A(\varphi)$ and $B(\varphi)$ of action (\ref{TeVeS}) also enter
the game. Moreover, one also needs to ensure that the matter
field equations always remain hyperbolic. We refer to
\cite{BEF07} for a discussion of these issues.

\subsection{Difficulties}
\label{subsec:6.3}
Although the class of relativistic models (\ref{TeVeS}) is the
most promising one to reproduce the MOND phenomenology, it has
anyway a long list of difficulties. Some of them can be solved,
but at the price of complicated and unnatural Lagrangians. For
instance, we mentioned above the problem of light deflection,
which is too small in conformally coupled scalar-tensor theories
(\ref{actionST}), therefore one needed to introduce a vector
field $U_\mu$ in action (\ref{TeVeS}). One may also notice that
such models are not very predictive, since it would have been
possible to predict fully different lensing and rotation curves.
One may thus consider them as fine-tuned fits rather than
fundamental theories imposed by some deep symmetry principles.
The most famous model, called TeVeS (for Tensor-Vector-Scalar)
\cite{Bekenstein:2004ne,Bekenstein:2004ca}, also presents some
discontinuous functions, and does not allow to pass smoothly from
a time evolution (cosmology) to a spatial dependence (local
physics in the vicinity of a galaxy). However, some cures are
possible \cite{BEF07}, although they again involve rather
unnatural refinements.

One peculiarity of the TeVeS model is that its author imposed
that gravitons and scalar particles are slower than photons
(\textit{a priori} to avoid causal paradoxes due to superluminal
propagation, assuming that light is a more fundamental field than
gravity). But Refs.~\cite{Moore:2001bv,Elliott:2005va} proved
that in such a case, high-energy cosmic rays would rapidly lose
their energy by Cherenkov radiation of gravitational waves, and
this would be inconsistent with their observation on Earth.
However, a very simple solution exists to cure this problem: One
just needs to flip a sign in one of the terms of the TeVeS model
\cite{Bekenstein:2004ne,Bekenstein:2004ca}, and merely accept
that photons can be slower than gravitons \cite{BEF07}. If all
the field equations are ensured to remain hyperbolic, with a
common time direction, no causal paradox can be caused by such a
situation.

The vector $U_\mu$ of action (\ref{TeVeS}) is assumed to be
timelike in the TeVeS model, therefore there \textit{a priori}
exists a preferred frame in which it takes the value $U_\mu =
(1,0,0,0)$ \cite{Sanders:1996wk,Foster:2005dk}. References
\cite{Bekenstein:2004ne,Bekenstein:2004ca} argue that this can
anyway be consistent with the high-precision tests of local
Lorentz invariance of gravity in the solar system if this vector
is dynamical. However, Ref.~\cite{Clayton:2001vy} showed that the
corresponding Hamiltonian is not bounded by below, precisely
because of the kinetic term of this vector field. Therefore, the
model is unstable. Other instabilities are also present in the
slightly different model of Ref.~\cite{Sanders:2005vd}.

A final difficulty is related to the post-Newtonian tests of
relativistic gravity, summarized in Sec.~\ref{sec:4} above. In
standard scalar-tensor theories (\ref{actionST}), Fig.~\ref{fig7}
shows that the linear matter-scalar coupling constant $\alpha_0$
should be small. In the TeVeS model, of the form (\ref{TeVeS}),
the functions $A(\varphi)$ and $B(\varphi)$ have been
\textit{tuned} to mimic the Schwarzschild metric of GR up to the
first post-Newtonian order, even for a large matter-scalar
coupling constant $\alpha_0$. Therefore, the full plane
$(|\alpha_0|,\beta_0)$ of Fig.~\ref{fig7} seems now allowed by
experimental data. However, binary-pulsar tests do not depend on
$A(\varphi)$ and $B(\varphi)$ in the same way, and one actually
gets basically the same constraints as in Figs.~\ref{fig10} and
\ref{fig11}, notably the tight bounds on $\alpha_0$ imposed by
the pulsar-white dwarf binary PSR J1141$-$6545 of
Fig.~\ref{fig11}. In conclusion, in spite of the tuning of
$A(\varphi)$ and $B(\varphi)$ to mimic GR in the solar system,
binary pulsars anyway impose that matter should be
\textit{weakly} coupled to the scalar field \cite{BEF07}. This is
quite problematic because we wish the same scalar field to give
rise, at large distances, to the MOND acceleration
$\sqrt{GMa_0}/r$ whose magnitude is fixed.
\begin{figure}[b]
\sidecaption
\includegraphics[scale=.37]{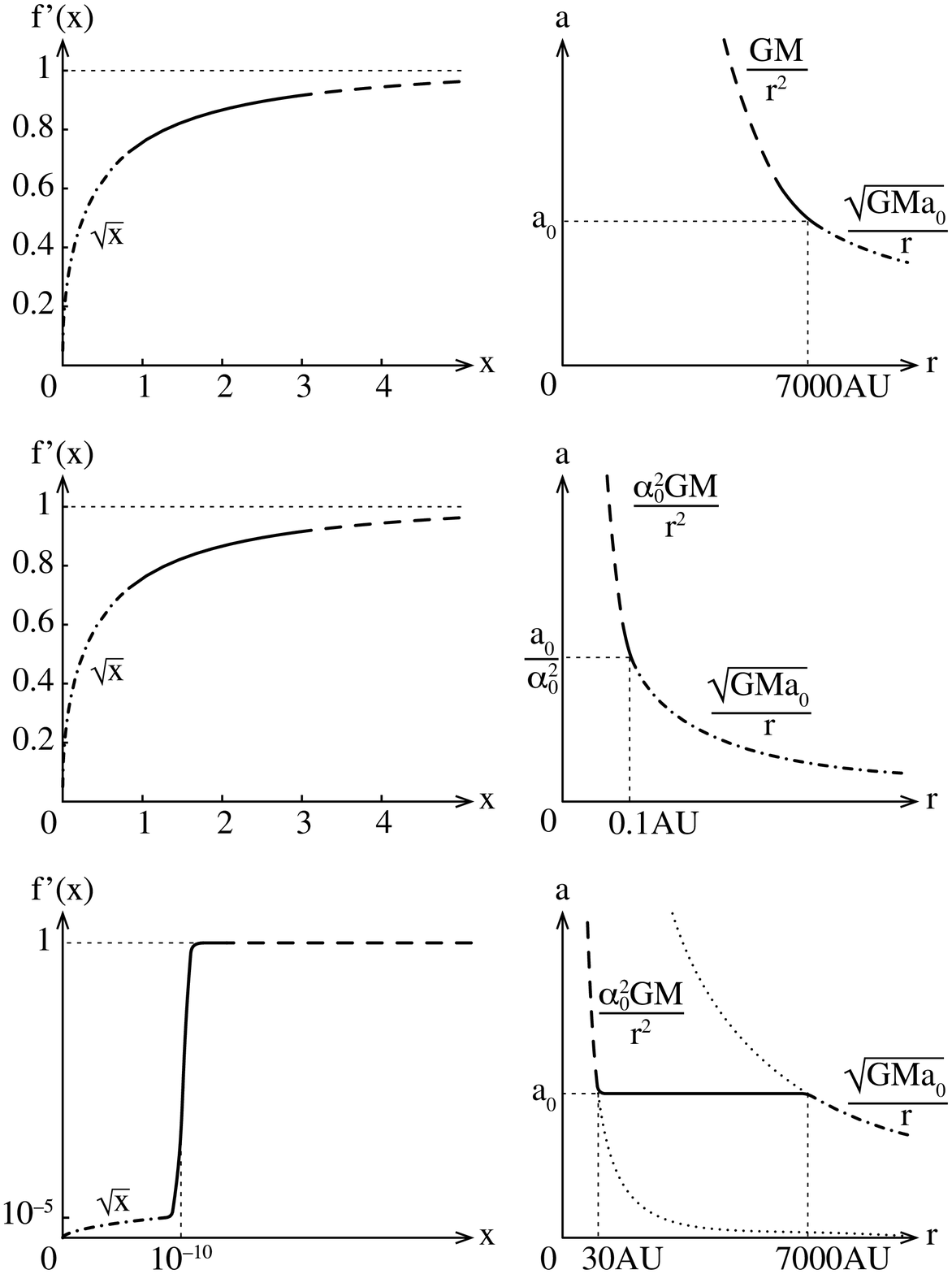}
\caption{Derivative of the function $f$ defining the kinetic term
of the scalar field in action (\ref{TeVeS}), such that the MOND
dynamics is predicted at large distances and Newtonian physics at
small distances. In the first case, $f'$ has a natural shape and
the MOND acceleration is obtained at the expected distance
$\sqrt{GM/a_0}$, but the scalar-field contribution to the
acceleration is too large to be consistent with binary-pulsar
tests. In the second case, $f'$ has the same natural shape and
the force caused by the scalar field is small enough when $r
\rightarrow 0$, but all planets should undergo an extra MOND
acceleration $\sqrt{GMa_0}/r$. In the third case, the scalar
force is small enough in the solar system and the MOND law is
predicted beyond the characteristic distance $\sqrt{GM/a_0}$, but
the shape of $f'$ is extremely unnatural.}
\label{fig12}
\end{figure}
Figure~\ref{fig12} illustrates schematically the difficulty (more
precise discussions are given in Ref.~\cite{BEF07}): If we wish
the function $f\left((\partial_\mu \varphi)^2\right)$ entering
action (\ref{TeVeS}) to have a natural enough shape, either
scalar-field effects are too large to be consistent with
binary-pulsar tests, or the solar-system size is already large
enough for all planets to be in the MOND regime and undergo a
small $\sqrt{GMa_0}/r$ acceleration in addition to the Newtonian
$G M/r^2$. This would be ruled out by tests of Kepler's third law
\cite{Talmadge:1988qz}. The only remaining solution would be to
impose a small enough $\alpha_0^2 GM/r^2$ acceleration caused by
the scalar field within the solar system, and the MOND law
$\sqrt{GMa_0}/r$ at large distances, but this would correspond to
the quite unnatural shape of the function $f'$ displayed in the
third panel of Fig.~\ref{fig12}. Although this is not yet
excluded experimentally, it would suffice to improve
binary-pulsar constraints by one order of magnitude to rule out
this kind of fine-tuned model, because one would need a shape of
function $f'$ violating the consistency conditions
(\ref{conditions}).

\subsection{Nonmininal couplings}
\label{subsec:6.4}
Another possible way to modify gravity is inspired by the
behavior of extended bodies within GR itself, as illustrated in
Sec.~\ref{sec:5} above. One may assume that matter is not only
coupled to the metric but also nonminimally to its curvature. Let
us thus consider an action of the form
\begin{equation}
S = \frac{c^3}{16\pi G_*}\int d^4 x\sqrt{-g^*}\, R^*
+S_{\rm matter}\left[{\rm matter}; g_{\mu\nu}
\equiv f(g^*_{\mu\nu}, R^*_{\lambda\mu\nu\rho},
\nabla^*_\sigma R^*_{\lambda\mu\nu\rho}, \dots)\right],
\end{equation}
which is Lorentz-invariant and satisfies the Einstein equivalence
principle because all matter fields are coupled to the same
tensor $g_{\mu\nu}$ (although it is built from the spin-2 metric
$g^*_{\mu\nu}$ and its curvature tensor). An immediate bonus of
this class of theories is that they reduce to GR in vacuum,
therefore standard solutions for the metric $g^*_{\mu\nu}$ remain
valid, notably the Schwarzschild solution for spherically
symmetric configurations. The only difference with GR is that
matter is no longer minimally coupled to $g^*_{\mu\nu}$, and
therefore that its motion can be changed. Reference \cite{BEF07}
showed that it is \textit{a priori} possible to reproduce the
MOND dynamics within this framework. However, the same theorem by
Ostrogradski \cite{Ostrogradski,Woodard:2006nt} that we mentioned
in Sec.~\ref{sec:2} suffices to conclude that these models are
generically unstable, and this can indeed be checked explicitly.

This instability can fortunately be avoided by slightly
complicating the model in vacuum. Instead of pure GR, let us
assume a scalar-tensor theory in vacuum, with a negligible scalar
mass $m_\varphi$ at galaxy scale. One may thus define an action
\begin{eqnarray}
S &=& \frac{c^3}{16\pi G_*}\int d^4 x\sqrt{-g^*}
\left\{R^*- 2g_*^{\mu\nu}\partial_\mu\varphi\partial_\nu\varphi
-2 m_\varphi^2 \varphi^2\right\}\nonumber\\
&&+S_{\rm matter}\left[{\rm matter}; g_{\mu\nu}
\equiv A^2[\varphi] g^*_{\mu\nu} + B[\varphi] \partial_\mu \varphi
\partial_\nu \varphi\right],
\label{nonMinScalar}
\end{eqnarray}
and show that specific functionals $A[\varphi] = A(\varphi,
\partial_\mu \varphi)$ and $B[\varphi]= B(\varphi, \partial_\mu
\varphi)$ allow us to reproduce both the MOND dynamics and the
right amount of light deflection by galaxies. The generic
Ostrogradskian instability is avoided because the matter action
involves second time-derivatives of the scalar field only
linearly \cite{BEF07}. Such a model is \textit{a priori} as good
as those which were previously proposed in the literature, and it
is much simpler to analyze. Because of its simplicity, it has
thus been possible to study its behavior \textit{within} matter
(an analysis which is usually too difficult to perform in other
models). The bad news is that the scalar field equations do not
remain hyperbolic within the dilute gas in outer regions of a
galaxy. Therefore, this class of models is finally also ruled out
to reproduce the MOND dynamics.

On the other hand, this class (\ref{nonMinScalar}) is able to
reproduce the Pioneer anomaly (if confirmed) in a consistent
relativistic theory of gravity, without spoiling the other
well-tested predictions of GR. This anomaly is actually a
\textit{simpler} problem than galaxy rotation curves. Indeed, one
of the difficulties of MOND models was to predict a force
$\propto \sqrt{M}$. In the case of the two Pioneer spacecrafts,
we do not know how their extra acceleration towards the Sun,
$\delta a \approx 8.5 \times 10^{-10}\, {\rm m.s}^{-2}$, is
related (or not) to the mass $M_\odot$ of the Sun. There are
thus actually several stable and well-posed models able to
reproduce this effect, that were constructed in Ref.~\cite{BEF07}
using the previous analyses of
\cite{Jaekel:2005qe,Jaekel:2005qz,Jaekel:2006me}. One of the
simplest models reads schematically
\begin{eqnarray}
S &=& \frac{c^3}{16\pi G_*}\int d^4 x\sqrt{-g^*}
\left\{R^*- 2g_*^{\mu\nu}\partial_\mu\varphi\partial_\nu\varphi
-2 m_\varphi^2 \varphi^2\right\}\nonumber\\
&&+S_{\rm matter}\left[{\rm matter}; g_{\mu\nu}
\equiv e^{2\alpha \varphi} g^*_{\mu\nu} -\lambda \frac{\partial_\mu \varphi
\partial_\nu \varphi}{\varphi^5}\right],
\label{nonMinPioneer}
\end{eqnarray}
where one needs to impose $\alpha^2 < 10^{-5}$ to pass
solar-system and binary-pulsar tests, $\lambda \approx
(\alpha G M_\odot/c^2)^3 (\delta a/v^2) \approx \alpha^3
(10^{-4}{\rm m})^2$ to \textit{fit} the Pioneer anomaly, and
where the scalar mass $m_\varphi$ needs to be negligibly small at
solar-system scale. Actually, refinements are necessary to define
correctly this model when $\varphi \rightarrow 0$, but we refer
to \cite{BEF07} for this discussion.

\section{Conclusions}
\label{sec:7}
In these lecture notes, we have underlined that the study of
contrasting alternatives to GR is useful to understand better
which features of the theory have been tested, and to suggest new
possible tests. One may modify either inertia (i.e., the matter
action $S_{\rm matter}$), gravity (i.e., the action defining its
dynamics, $S_{\rm gravity}$), or consider nonminimal couplings to
curvature. We stressed that the dynamics of gravity directly
influences the motion of massive bodies, and this was
illustrated by constructing the Fokker action.

Our study of scalar-tensor theories of gravity exhibited a
\textit{qualitative} difference between solar-system experiments
(weak fields) and binary-pulsar tests (strong fields). While the
former tightly constrain the linear matter-scalar coupling
constant, the latter are nonperturbatively sensitive to nonlinear
couplings, and notably forbid that the quadratic matter-scalar
coupling constant be large and negative. We mentioned that the
no-hair theorem imposes that the motion of black holes in
scalar-tensor theories is the same as in GR, contrary to other
massive bodies. We also gave a simple dimensional argument
showing that finite-size effects are much larger in alternative
theories than in GR.

Finally, we illustrated that the MOND phenomenology may \textit{a
priori} be reproduced within any of the above classes of
alternative theories (modified inertia, modified gravity,
couplings to curvature), but that all the proposed models present
several experimental and theoretical difficulties. However, many
possible routes remain possible, and generalizations of
Einstein-aether theories
\cite{Jacobson:2000xp,Jacobson:2004ts,Eling:2004dk,JacobsonReport,
Zlosnik:2006zu} able to reproduce MOND deserve to be studied,
notably from the point of view of stability and well-posedness
of the Cauchy problem. Of course, the simplest solution to
account for galaxy rotation curves and cosmological data seems
merely to accept the existence of dark matter. Instead of
modifying gravity, one may then even devise some ``modified dark
matter'' so that it reproduces the MOND successes while keeping
the standard cold dark matter cosmological scenario
\cite{Blanchet:2006yt,BlanchetLeTiec1,BlanchetLeTiec2}.

\end{document}